\documentclass[twocolumn]{aastex631}

\usepackage{xcolor}
\definecolor{MM}{HTML}{BF16ED}
\usepackage{subfigure}

\submitjournal{The Astronomical Journal}

\shorttitle{\textit{HST} Emission Spectrum of WASP-77Ab}
\shortauthors{Mansfield et al.}

\begin{document}

\title{Confirmation of Water Absorption in the Thermal Emission Spectrum of the Hot Jupiter WASP-77Ab with HST/WFC3}

\correspondingauthor{Megan Mansfield}
\email{meganmansfield@arizona.edu}

\author[0000-0003-4241-7413]{Megan Mansfield}
\affiliation{Steward Observatory, University of Arizona, 933 N. Cherry Ave., Tucson, AZ 85719, USA}
\affiliation{NHFP Sagan Fellow}

\author[0000-0002-3295-1279]{Lindsey Wiser}
\affiliation{School of Earth and Space Exploration, Arizona State University, Tempe, AZ 85281, USA}

\author[0000-0002-7352-7941]{Kevin B. Stevenson}
\affiliation{Johns Hopkins University Applied Physics Laboratory, Laurel, MD 20723, USA}

\author{Peter Smith}
\affiliation{School of Earth and Space Exploration, Arizona State University, Tempe, AZ 85281, USA}

\author{Michael R. Line}
\affiliation{School of Earth and Space Exploration, Arizona State University, Tempe, AZ 85281, USA}

\author[0000-0003-4733-6532]{Jacob L.\ Bean}
\affiliation{Department of Astronomy \& Astrophysics, University of Chicago, Chicago, IL 60637, USA}

\author{Jonathan J. Fortney}
\affiliation{Department of Astronomy and Astrophysics, University of California, Santa Cruz, CA 95064, USA}

\author{Vivien Parmentier}
\affiliation{Department of Physics, University of Oxford, Oxford, OX1 3PU, UK}

\author{Eliza M.-R. Kempton}
\affiliation{Department of Astronomy, University of Maryland, College Park, MD 20742, USA}

\author{Jacob Arcangeli}
\affiliation{Anton Pannekoek Institute for Astronomy, University of Amsterdam, 1098 XH Amsterdam, The Netherlands}

\author[0000-0002-0875-8401]{Jean-Michel D\'esert}
\affiliation{Anton Pannekoek Institute for Astronomy, University of Amsterdam, 1098 XH Amsterdam, The Netherlands}

\author{Brian Kilpatrick}
\affiliation{Space Telescope Science Institute, Baltimore, MD 21218, USA}

\author{Laura Kreidberg}
\affiliation{Max Planck Institute for Astronomy, 69117 Heidelberg, Germany}

\author{Matej Malik}
\affiliation{Department of Astronomy, University of Maryland, College Park, MD 20742, USA}

\begin{abstract}

Secondary eclipse observations of hot Jupiters can reveal both their compositions and thermal structures. Previous observations have shown a diversity of hot Jupiter eclipse spectra, including absorption features, emission features, and featureless blackbody-like spectra. We present a secondary eclipse spectrum of the hot Jupiter WASP-77Ab observed between $1-5$~$\mu$m with the \textit{Hubble Space Telescope} (\textit{HST}) and the \textit{Spitzer Space Telescope}. The \textit{HST} observations show signs of water absorption indicative of a non-inverted thermal structure. We fit the data with both a one-dimensional free retrieval and a grid of one-dimensional self-consistent forward models to confirm this non-inverted structure. The free retrieval places a $3\sigma$ lower limit on the atmospheric water abundance of $\log(n_\mathrm{H_2O})>-4.78$ and can not constrain the CO abundance. The grid fit produces a slightly super-stellar metallicity and constrains the carbon-to-oxygen ratio to less than or equal to the solar value. We also compare our data to recent high-resolution observations of WASP-77Ab taken with the Gemini-South/IGRINS spectrograph and find that our observations are consistent with the best-fit model to the high-resolution data. However, the metallicity derived from the IGRINS data is significantly lower than that derived from our self-consistent model fit. We find that this difference may be due to disequilibrium chemistry, and the varying results between the models applied here demonstrate the difficulty of constraining disequilibrium chemistry with low-resolution, low wavelength coverage data alone. Future work to combine observations from IGRINS, \textit{HST}, and \textit{JWST} will improve our estimate of the atmospheric composition of WASP-77Ab. 

\end{abstract}


\section{Introduction}

Thermal emission measurements taken during secondary eclipse have the potential to reveal information on both the compositions and thermal structures of hot Jupiter atmospheres. The compositions of hot Jupiter atmospheres can be used to track their formation and migration conditions \citep{Venturini2016,Madhusudhan2017}. For example, a key prediction of the core accretion theory of planet formation is that atmospheric metallicities should be inversely proportional to planet mass \citep{Fortney2013}. Furthermore, the carbon-to-oxygen (C/O) ratio provides information on the mechanisms through which hot Jupiters form and migrate to their current locations \citep{Oberg2011,Madhusudhan2014,Mordasini2016,AliDib2017,Espinoza2017,Schneider2021}.

In addition to constraining the composition, secondary eclipse observations can provide information on the thermal structures of hot Jupiters. Theory predicts a continuum of thermal structures and resulting secondary eclipse spectra, which can be divided into three primary categories \citep{Fortney2008,Parmentier2018}. The coolest hot Jupiters with dayside temperatures ($T_{day}$) below $\approx2100$~K are predicted to have non-inverted temperature-pressure (T-P) profiles, which cause absorption features in their emergent spectra. Hot Jupiters with intermediate temperatures between $2100<T_{day}<2400$~K should have emission features resulting from inverted T-P profiles. Such thermal inversions are predicted to be driven by the presence of a variety of chemical species, such as TiO, VO, FeH, and metal atoms \citep{Hubeny2003,Lothringer2018}. Finally, the ultra-hot Jupiters with $T_{day}>2400$~K are expected to also have strongly inverted T-P profiles, but display featureless secondary eclipse spectra in the \textit{HST}/WFC3 bandpass ($1.1-1.7$~$\mu$m) due to molecular dissociation and H$^{-}$ opacity \citep{Parmentier2018,Lothringer2018,Kitzmann2018}.

These predictions have been borne out through \textit{HST} observations of absorption features in low-temperature hot Jupiters (e.g., WASP-43b, \citealp{Kreidberg2014}; and HD\,209458b, \citealp{Line2016}), subtle emission features in medium-temperature hot Jupiters (e.g., WASP-121b, \citealp{evans17,MikalEvans2020,Mansfield2021}; and WASP-76b, \citealp{Edwards2020,Fu2020,Mansfield2021}) and blackbody-like spectra in the highest-temperature ultra-hot Jupiters (e.g., WASP-18b, \citealp{Arcangeli2018}; and WASP-103b, \citealp{Kreidberg2018}). However, not all observed hot Jupiters fit neatly into these three categories. For example, ultra-hot Jupiter Kepler-13Ab shows absorption features indicative of a non-inverted atmosphere, despite having a high dayside temperature of $\approx3000$~K \citep{Beatty2017}. In general, the population of observed planets shows a scatter in the water feature strengths at a given temperature, which may be caused by variations in atmospheric composition \citep{Mansfield2021}.

In this paper we present the secondary eclipse spectrum of WASP-77Ab observed with \textit{HST}/WFC3 between $1.1-1.7$~$\mu$m and \textit{Spitzer}/IRAC at 3.6 and 4.5~$\mu$m. WASP-77Ab is a mid-temperature hot Jupiter with an equilibrium temperature of $T_{eq}=1705$~K \citep{Maxted2013}, which is near the point where models predict a transition from non-inverted T-P profiles creating absorption features to inverted T-P profiles creating emission features \citep{Mansfield2021}. The exact temperature of this transition, however, depends in detail on parameters such as the planet's atmospheric composition and the amount of heat deposited in its interior. Our observations of WASP-77Ab have double the signal-to-noise of any previous observations at temperatures near this transition, giving us an opportunity to constrain the nature of this transition. We describe our observations and data reduction in Section~\ref{sec:observe}. In Section~\ref{sec:models}, we perform a 1D free retrieval on our data and compare our data to a set of 1D radiative-convective-thermochemical equilibrium models. Finally, in Section~\ref{sec:discuss} we compare our data to a recent Gemini-S/IGRINS high-resolution thermal emission spectrum of WASP-77Ab, compare the water feature strength of WASP-77Ab to the broader population, and discuss the results of our model fits.

\section{Observations and Data Reduction}
\label{sec:observe}

All of the data presented in this paper were obtained from the Mikulski Archive for Space Telescopes (MAST) at the Space Telescope Science Institute. The specific observations analyzed can be accessed via \dataset[10.17909/gjbj-r870]{https://doi.org/10.17909/gjbj-r870}.

\subsection{\textit{HST}/WFC3 Data}

We observed two secondary eclipses of WASP-77Ab on 2020 November 7 and 2020 December 19 using the \textit{HST}/WFC3+G141 grism between 1.1 and 1.7~$\mu$m as part of program GO-16168. Each visit consisted of five consecutive orbits in which WASP-77Ab was visible for approximately 52 minutes per orbit. At the beginning of each orbit, we took a direct image of the target with the F126N filter for wavelength calibration.

The observations were taken in the spatial scan mode with the $256\times256$ subarray using the SPARS25, NSAMP $=5$ readout pattern, resulting in an exposure time of 89.662~s. We used a scan rate of 0.195\,arcsec\,s$^{-1}$, which produced spectra extending approximately 153 pixels in the spatial direction and peak pixel counts of $\approx37,000$ electrons per pixel. We used bidirectional scans and observed 18 exposures per orbit.

We reduced the data using the data reduction pipeline described in \citet{Kreidberg2014a}. We used an optimal extraction procedure \citep{Horne1986} and masked cosmic rays. To subtract the background out of each frame, we visually inspected the images to find a clear background spot on the detector and subtracted the median of this background area. The uncertainties on the measurements were determined by adding in quadrature the photon noise, read noise, and median absolute deviation of the background.

\begin{figure}
    \centering
    \includegraphics[width=\linewidth]{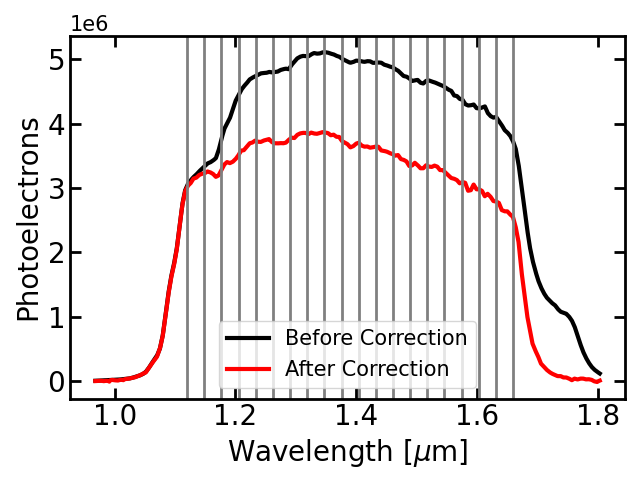}
    \caption{Example stellar spectrum extracted from one spatially scanned exposure taken by \textit{HST}/WFC3. Black and red lines indicate the extracted flux before and after correcting for the flux of the companion star, WASP-77B, respectively. The spectrum of WASP-77B appears redder in wavelength than that of WASP-77A because its spectral trace was slightly offset in the spectral direction on the detector. Vertical gray lines indicate the extent of the bins for the spectroscopic light curve.}
    \label{fig:obsspec}
\end{figure}

Following standard procedure for \textit{HST}/WFC3 eclipse observations, we discarded the first orbit of each visit. The spectra were binned into 19 channels at a resolution $R\approx40-60$. Figure~\ref{fig:obsspec} shows an example extracted stellar spectrum with the wavelength bins indicated. We also created a broadband white light curve by summing the spectra over the entire wavelength range.

We fit both the white light curve and spectroscopic light curves with the model described in \citet{Kreidberg2014a}, which includes an eclipse model \citep{Kreidberg2015} and a systematics model based on \citet{Berta2012}. For the white light curves, the free parameters in the eclipse model were the mid-eclipse time $T_{0}$ and the planet-to-star flux ratio $F_{p}$/$F_{s}$. For the spectroscopic light curves, the mid-eclipse time was fixed to the best-fit value from the white light curve ($T_{sec}=2455871.12983^{+0.00051}_{-0.00050}$~BJD$_{\mathrm{TDB}}$) and the only free parameter in the eclipse model was $F_{p}$/$F_{s}$. In both cases, the period, eccentricity, ratio of the semi-major axis to the stellar radius, inclination, and planet-to-star radius ratio were fixed to $P=1.360030$, $e=0$, $\frac{a}{R_{*}}=5.43$, $i=89.40$, and $ \frac{R_{p}}{R_{*}}=0.13012$, respectively \citep{Stassun2017,Turner2016}. The instrument systematics model included an orbit-long ramp, whose amplitude and offset were fixed to the same value for both visits, and a normalization constant, visit-long slope, and correction for an offset between scan directions, which all varied between visits. The white light curve fit thus contained a total of 10 free parameters, while the spectroscopic light curve fits had 9 free parameters.

WASP-77A has a companion star, WASP-77B, which has a projected distance large enough that their spectra do not overlap in stare mode. However, the spectra of these two stars overlap during spatial scans. In order to correct for this overlap, we observed a single 0.556\,s stare mode exposure with the G141 grism at the beginning of each of the two visits. For each visit, we used the same optimal extraction procedure \citep{Horne1986} to extract the stare mode spectra of WASP-77A and WASP-77B. We then corrected the observed flux for the presence of the companion star using the equation
\begin{equation}
    F_{*,corr}=F_{*,obs}\left(\frac{F_{A}}{F_{A}+F_{B}}\right),
\end{equation}
where $F_{*,corr}$ is the corrected flux in units of electrons, $F_{*,obs}$ is the observed flux in units of electrons, and $F_{A}$ and $F_{B}$ are the observed fluxes of the primary and companion star in that bandpass, respectively.

\begin{figure*}
    \begin{minipage}[c]{0.55\textwidth}
    \includegraphics[width=\linewidth]{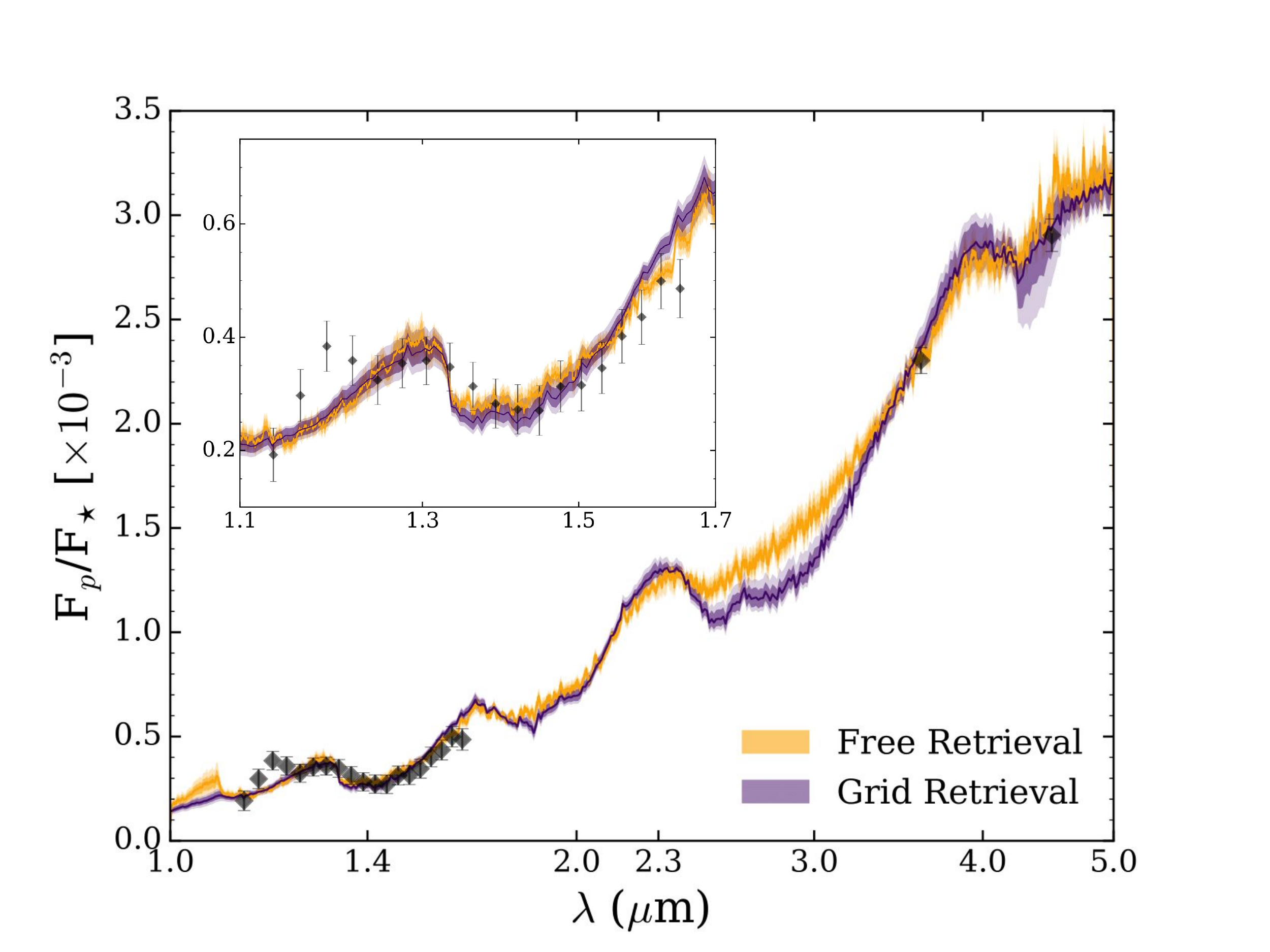}
    \end{minipage}
\hspace{1 mm}
\begin{minipage}[c]{0.4\textwidth}
    \includegraphics[width=\linewidth]{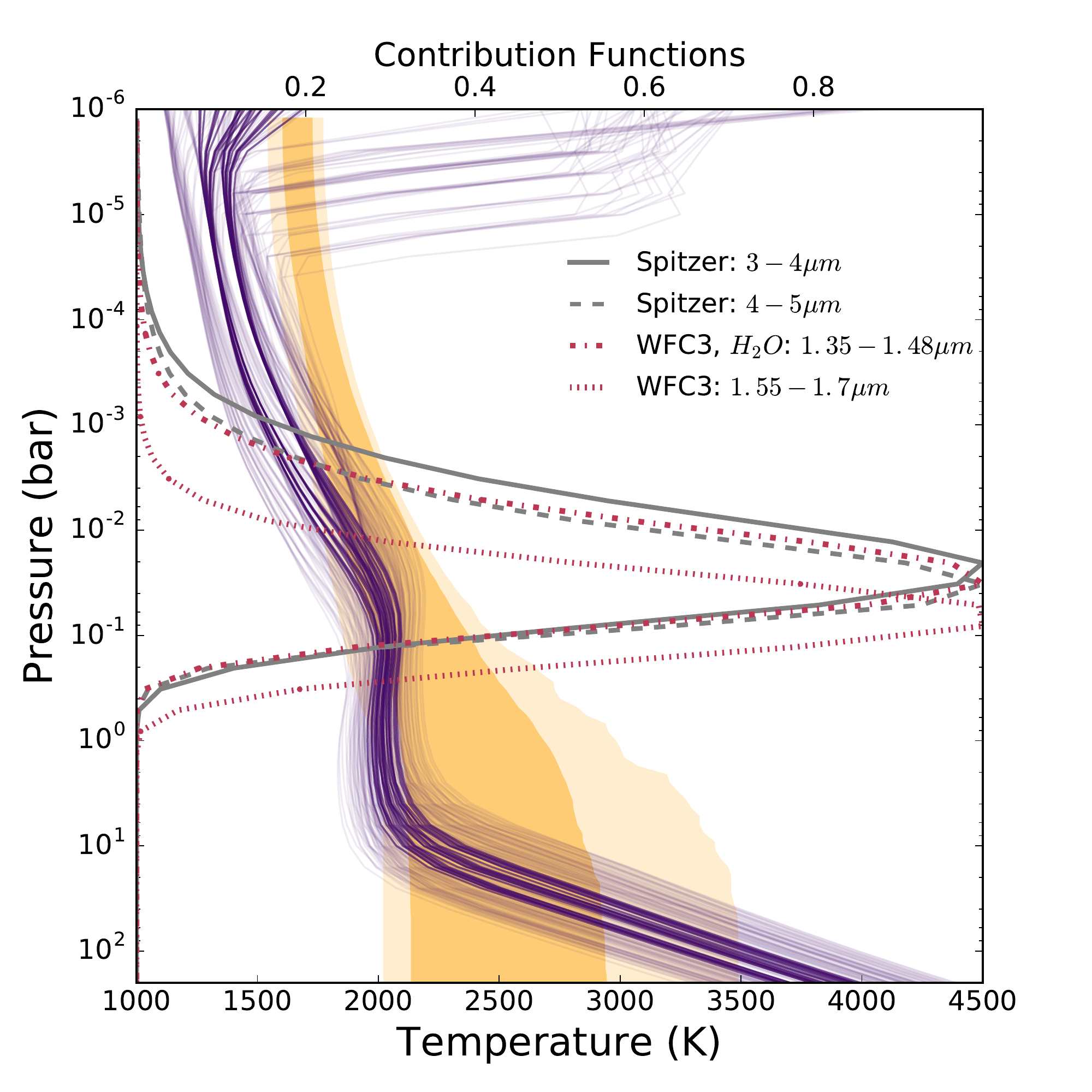}
    \end{minipage}
    \caption{\textbf{Left:} Emission spectrum fits using the 1D free retrieval described in Section ~\ref{sec:1dfree}, and the Sc-CHIMERA 1D model grid described in Section~\ref{sec:1dmodel}. Dark lines represent the median fit, and dark and light shading show 1 and 2$\sigma$ regions, respectively. Black points show the observations. The inset shows a zoomed in view of the WFC3 segment of the spectrum. The models generally fit the data well, with the best fit models having reduced chi squared of $\chi^{2}_{\nu}=1.12$ for the free retrieval and $1.24$ for the grid fit. \textbf{Right:} Corresponding pressure-temperature profiles. Grid profiles within 1 and 2$\sigma$ are shown by dark and light purple lines, respectively. 1 and 2$\sigma$ regions for the free retrieval are shaded in dark and light yellow, respectively. Contribution functions are also plotted for each \textit{Spitzer} point, and for in ($1.35-1.48$~$\mu$m) and out ($1.55-1.7$~$\mu$m) of the water feature in the WFC3 wavelength range.}
    \label{fig:1dspec}
\end{figure*}

We estimated the parameters with a Markov Chain Monte Carlo (MCMC) fit using the \texttt{emcee} package \citep{Foreman2013}. The best-fit white light curve had $\chi^{2}_{\nu}=4.92$ and an average residual of 90~ppm, which is typical for WFC3 observations of transiting planets orbiting bright host stars. The spectroscopic light curves achieved photon-limited precision, with $\chi^{2}_{\nu}$ values between $0.68-1.30$. The final secondary eclipse spectrum is shown in Figure~\ref{fig:1dspec}, and Table~\ref{tab:spectrum} lists the planet-to-star flux ratio in each channel.

\begin{table}[]
    \centering
    \begin{tabular}{c | c}
    \hline
    Wavelength [$\mu$m] & $F_{p}$/$F_{s}$ [ppm] \\
    \hline \hline
    $1.120-1.148$ & $192 \pm 47$ \\
    $1.148-1.177$ & $297 \pm 46$ \\
    $1.177-1.205$ & $384 \pm 45$ \\
    $1.205-1.234$ & $359 \pm 44$ \\
    $1.234-1.262$ & $324 \pm 43$ \\
    $1.262-1.291$ & $354 \pm 43$ \\
    $1.291-1.319$ & $359 \pm 42$ \\
    $1.319-1.347$ & $348 \pm 42$ \\
    $1.347-1.376$ & $313 \pm 43$ \\
    $1.376-1.404$ & $283 \pm 43$ \\
    $1.404-1.433$ & $273 \pm 44$ \\
    $1.433-1.461$ & $271 \pm 44$ \\
    $1.461-1.489$ & $313 \pm 45$ \\
    $1.489-1.518$ & $315 \pm 46$ \\
    $1.518-1.546$ & $346 \pm 45$ \\
    $1.546-1.575$ & $402 \pm 47$ \\
    $1.575-1.603$ & $436 \pm 48$ \\
    $1.603-1.632$ & $499 \pm 49$ \\
    $1.632-1.660$ & $486 \pm 51$ \\
    \hline
    $3.6$ & $2303 \pm 62$ \\
    $4.5$ & $2904 \pm 78$ \\
    \hline
    \end{tabular}
    \caption{Secondary eclipse spectrum of WASP-77Ab.}
    \label{tab:spectrum}
\end{table}

\subsection{\textit{Spitzer}/IRAC Data}


The Spitzer Space Telescope observed the WASP-77 system at 3.6 and 4.5 {\micron} under program 13038 (PI: Stevenson).  Each phase curve observation lasted 39.5 hours (starting shortly before secondary eclipse and ending shortly after the subsequent eclipse) and was subdivided into three Astronomical Observation Requests (AORs).  The first AOR consisted of a 24-minute settling period, followed by a two science AORs lasting 23 and 16 hours each.  The break between science AORs occurred shortly after transit.

We used the Photometry for Orbits, Eclipses, and Transits (POET) data reduction and analysis pipeline \citep{Stevenson2012a,Cubillos2013,Bell2021} to derive the secondary eclipse depths reported in this work.  
For these data, we utilized a $3\times3$-pixel centroiding aperture to minimize contamination from WASP-77A's nearby binary companion (WASP-77B) located roughly 2.5 Spitzer pixels away.  The standard $5\times5$-pixel centroiding aperture demonstrated a noticeable bias towards WASP-77B and significant volatility in the measured values.  
At 3.6 {\micron}, the pointing was stable over the course of the phase curve observation.  At 4.5 {\micron}, we measured a drift of 0.5 pixels over the first six hours of observing before stabilizing.  The 4.5 {\micron} centroids do not overlap with the ``sweet spot'' mapped out by \citet{May2020} and, thus, we could not use their fixed intrapixel sensitivity map to remove position-dependent systematics.

We tested a range of photometry aperture sizes from 2.0 to 4.75 pixels in 0.25-pixel increments.  For each aperture size, we fit the transit, two eclipses, and sinusoidal variation from the planet.  Both Spitzer channels use BLISS mapping \citep{Stevenson2012a} to fit the intrapixel sensitivity variations.  The 3.6 {\micron} observation also requires a rising exponential plus linear ramp to fit the time-dependent systematics and a linear function to fit variations in PRF width along the $y$ direction \citep[PRF detrending,][]{Lanotte2014}.  The 4.5 {\micron} channel does not exhibit a time-dependent systematic.

The measured eclipse depths decrease systematically with increasing photometry aperture size due to increasing contamination from WASP-77B within the aperture. We use the mean image of each Spitzer observation to estimate the companion flux fraction within each photometric aperture.  This process involves masking the flux from WASP-77A, computing the centroid of WASP-77B, and performing aperture photometry on a Spitzer PRF situated at WASP-77B's centroid position.  We then follow the methods describe by \citet{Stevenson2014a} to compute corrected eclipse depths.  Using CatWISE \citep{Marocco2021}, we estimate the dilution factor to be $0.410{\pm}0.013$ and $0.405{\pm}0.012$ at 3.6 and 4.5 {\micron}, respectively.  This calculation is possible since WISE1 and WISE2 have similar bandpasses to IRAC1 and IRAC2.  As validation to our methods, we find that the corrected eclipse depths are independent of our choice of aperture size (i.e., they are all consistent within 1$\sigma$).  Using the apertures that yield the smallest standard deviation of the normalized residuals (3.5 pixels at 3.6 {\micron} and 4.5 pixels at 4.5 {\micron}), we report our final eclipse depths in Table~\ref{tab:spectrum}.

\section{Analysis} 
\label{sec:models}

We explore fitting the data with a variety of models to test how a gradient of model assumptions impact the derived atmospheric parameters. 
Here we explore the results from two common modeling philosophies. The first, described in Section~\ref{sec:1dfree}, is the "free" retrieval methodology whereby we fit for the constant-with-altitude abundances for water and carbon monoxide, (the dominant species over the observed wavelengths) and a vertical temperature profile. Within the free retrieval there are no physical/chemical constraints that relate the gas abundances to each other or the temperature profile.  The second, described in Section~\ref{sec:1dmodel}, is the self-consistent 1D radiative convective grid model fitting method. This method assumes thermochemical equilibrium chemical abundances for all gases along the temperature pressure profile, which in turn is dependent upon the opacities and gas abundances.  In this framework, rather than retrieving the gas abundances and T-P profile independently, we instead retrieve intrinsic elemental abundances (parameterized with a metalllicity and carbon to oxygen ratio) and a heat redistribution (which sets the effective stellar flux on the planetary dayside). 
We explore both of these models throughout this paper because their differing levels of complexity allow us to better understand the nature of the planet's atmosphere than applying a single model framework alone.

\subsection{1D Free Retrieval}
\label{sec:1dfree}

We performed a 9-parameter free atmospheric retrieval, fitting directly for the volume mixing ratios (constant with pressure) of H$_2$O and CO, 6 parameters describing the shape of an analytic temperature-pressure profile, and a scale factor (see Table \ref{tab:free priors} for each model parameter and its prior range, which is uniform for all parameters). The scale factor ($a$) accounts for any geometric dilution of a dayside hotspot by multiplying the planet-to-star flux ratio by a constant \citep[e.g.,][]{Taylor2020}. A value of $a$ close to 1 indicates a more homogeneous dayside, while a smaller value of $a$ indicates a more concentrated hotspot. The temperature-pressure profile is that given by \citet{madhu2009}, which is a piecewise function of the form

\begin{equation}
    T(P) = T_0 + \Bigg( \frac{\log(P/P_0)}{\alpha_1} \Bigg)^{1/\beta_1} , \ P_0 < P < P_1
    \label{eqn:layer1}
\end{equation}

\begin{equation}
    T(P) = T_2 + \Bigg( \frac{\log(P/P_2)}{\alpha_2} \Bigg)^{1 / \beta_2} , \ P_2 < P < P_3
    \label{eqn:layer2}
\end{equation}

\begin{equation}
    T(P) = T_3 , \ P > P_3
    \label{eqn:layer3}
\end{equation}
for three atmospheric layers. Layer 1, the upper atmosphere, is between pressures $P_0$ (the top of atmosphere) and $P_1$, and the T-P profile has a slope determined by $\alpha_1$ and $\beta_1$. Layer 2, the middle atmosphere, is between pressures $P_1$ and $P_3$ and has a slope determined by $\alpha_2$ and $\beta_2$. At pressures greater than $P_3$, the profile is isothermal. A third pressure point, $P_2$, can be either above or below $P_1$, and if $P_2 > P_1$, an inversion will occur. $T_2$ and $T_3$ are the temperatures at $P_2$ and $P_3$ (determined via continuity), respectively, and $T_0$ is the top-of-atmosphere temperature. We set $\beta_1 = \beta_2 = 0.5$ to match empirical results, and are left with 6 free parameters: $T_0$, $P_1$, $P_2$, $P_3$, $\alpha_1$, and $\alpha_2$. While an inversion is not expected for WASP-77Ab, we allow $P_2$ to range both higher and lower than $P_1$.

For the planet's thermal emission spectrum, we use a psuedo-line-by-line radiative transfer code with absorption cross sections sampled at a resolution $R = \lambda / \Delta \lambda$ of 20,000 \citep[for an introduction to the forward modelling and retrieval frameworks, see][]{Line2013, Line2021}. We only include opacities of H$_2$-H$_2$/He CIA, H$_2$O, and CO. The stellar spectrum is interpolated from the \texttt{PHOENIX} library of model stellar spectra \citep{Husser2013} and smoothed with a Guassian filter. Each model spectrum is then binned onto the WFC3 wavelength grid and integrated through the IRAC throughput curves. We used the Python wrapper \texttt{PyMultiNest} \citep{pymultinest} for the nested sampling algorithm \texttt{MULTINEST} \citep{multinest} for Bayesian parameter estimation. 

\begin{table}[]
    \centering
    \begin{tabular}{c|c}
    \hline
        Parameter & Prior \\
        \hline 
        \hline
        $\log(n_\mathrm{H_2O})$ & $\mathcal{U}$(-12, 0) \\
        $\log(n_\mathrm{CO})$ & $\mathcal{U}$(-12, 0) \\
        $T_0$ [K] & $\mathcal{U}$(400,3000) \\
        $\log P_1 [\log$ bar] & $\mathcal{U}$(-5.5, 2.5) \\
        $\log P_2 [\log$ bar] & $\mathcal{U}$(-5.5, 2.5) \\
        $\log P_3 [\log$ bar] & $\mathcal{U}$(-2, 2.5) \\
        $\alpha_1$ & $\mathcal{U}$(0.02, 1.98) \\
        $\alpha_2$ & $\mathcal{U}$(0.02, 1.98) \\
        $\log(a)$ & $\mathcal{U}$(-2, 4) \\
    \hline
    \end{tabular}
    \caption{Free parameters and their prior ranges (all uniform) for the free retrieval. }
    \label{tab:free priors}
\end{table}

\begin{figure*}[!ht]
    \centering
    \includegraphics[width=\textwidth]{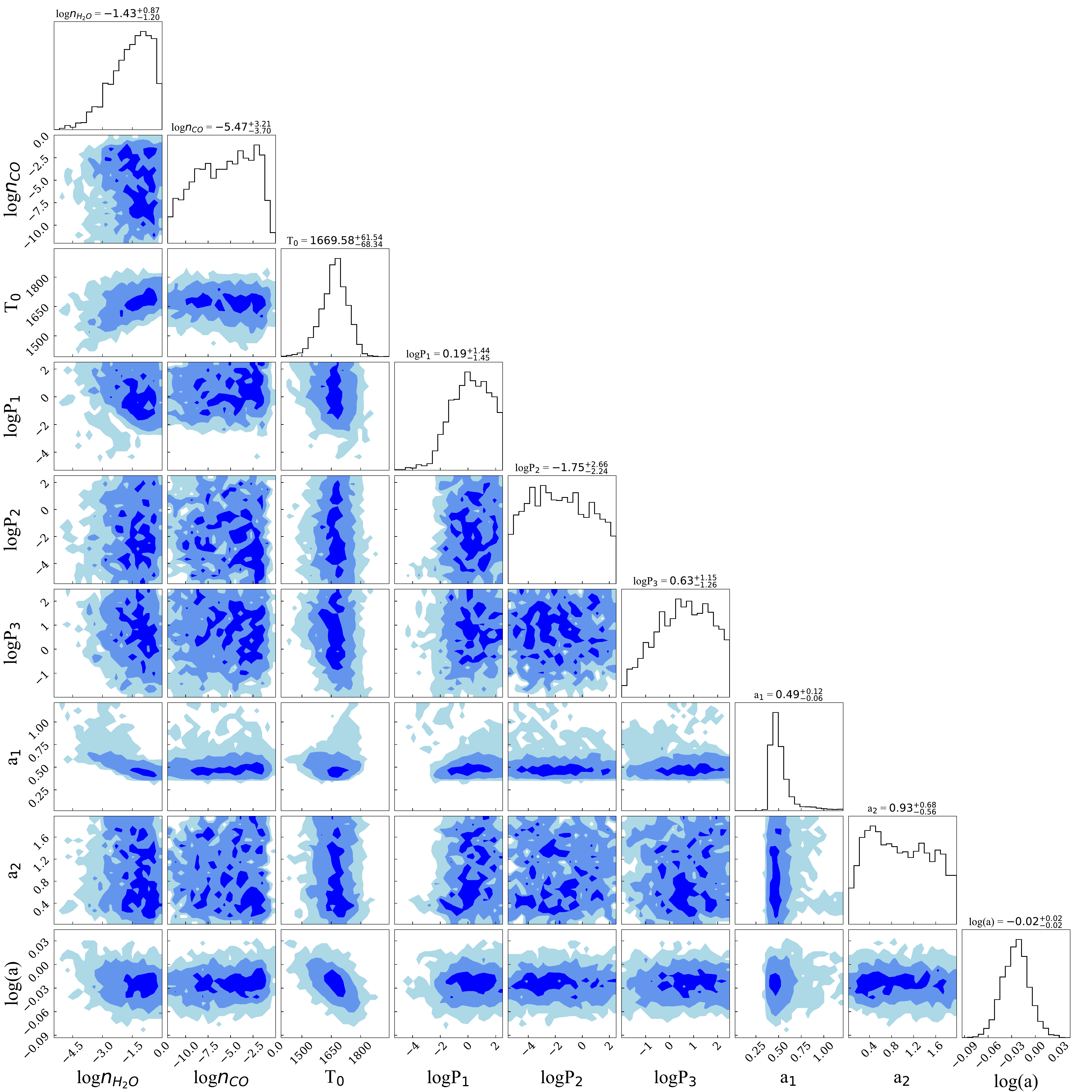}
    \caption{Posterior distributions of all parameters in the free retrieval (Section \ref{sec:1dfree}). Off-diagonal plots show 2D posterior probabilities for pairs of parameters, with 1, 2, and $3\sigma$ intervals indicated in dark, medium, and light blue. Panels on the diagonal show 1D posterior probability distributions for each parameter. The free parameters include volume mixing ratios of H$_2$O and CO, 6 parameters for the analytic T-P profile, and a scale factor.}
    \label{fig:full corner}
\end{figure*}

Figure \ref{fig:1dspec} shows the measured spectrum for WASP-77Ab with model spectra and T-P profiles randomly drawn from the posterior distribution. Figure \ref{fig:full corner} contains a corner plot showing the marginal posterior probability distribution of each parameter. The best fit spectrum has a reduced chi-square metric $\chi^2_\nu$ = 1.12. We are unable to constrain the abundance of H$_2$O but can place a robust lower limit at $\log(n_\mathrm{H_2O})>-4.78$ at 3$\sigma$, whereas CO (and therefore C/O) is entirely unconstrained due to the lack of significant CO spectral features captured by WFC3. Consequently, we can only place a lower limit on the metal content of the atmosphere at [(C+O)/H] $>$ -1.69 at 3$\sigma$. The retrieved temperature-pressure profile (Figure~\ref{fig:1dspec}, right) is monotonically increasing with pressure and has a top-of-atmosphere temperature of 1670$^{+62}_{-68}$ K. The scale factor is 0.95 $\pm$ 0.04, indicative of a homogenous dayside with little to no clouds.

\subsection{Comparison to 1D Model Grid}
\label{sec:1dmodel}
In addition to performing a classic free retrieval, we used Sc-CHIMERA to perform a 1D radiative-convective-thermochemical equilibrium (1D-RC) grid model retrieval, following a similar methodology described in \cite{Arcangeli2018} and \cite{Mansfield2018}. We generated a WASP-77Ab specific model grid with free parameters for the global heat redistribution ($f$), metallicity ([M/H]), and carbon-to-oxygen ratio (C/O). We used the same 1D-RC framework described in \citet{Mansfield2021}, which is an upgrade to that used in \cite{Arcangeli2018} and \cite{Mansfield2018}. We also include the scale factor ($a$), which can be included without an additional grid dimension. The grid is coupled to the {\tt pymultinest} nested sampler to perform parameter estimation across $f$, [M/H], C/O, and $a$. 

We defined prior ranges of 0.4 -- 2.8 for $f$, -2.0 -- 2.6 for [M/H], and 0.01 -- 1.4 for C/O. The heat redistribution factor is defined as in \citet{Parmentier2020} as a function of dayside and equilibrium temperature, $f=(T_d/T_*)^{4}$. As such, $f=1$ corresponds to full redistribution, $f=2$ corresponds to dayside-only redistribution, and $f=2.67$ is the maximum value allowed by energy conservation. The prior range for $f$ extends beyond possible values to allow {\tt pymultinest} to converge close to maximum and minimum, if needed. The prior range for [M/H] encompasses the range of Solar System and exoplanet observations and predictions presented in prior literature \citep{Thorngren2016, Mordasini2016, Kreidberg2014, Welbanks2019}. The C/O prior range is also defined based on prior literature expectations of C/O$<1$ \citep{Mordasini2016}.

\begin{figure}
    \centering
    \includegraphics[width=\linewidth]{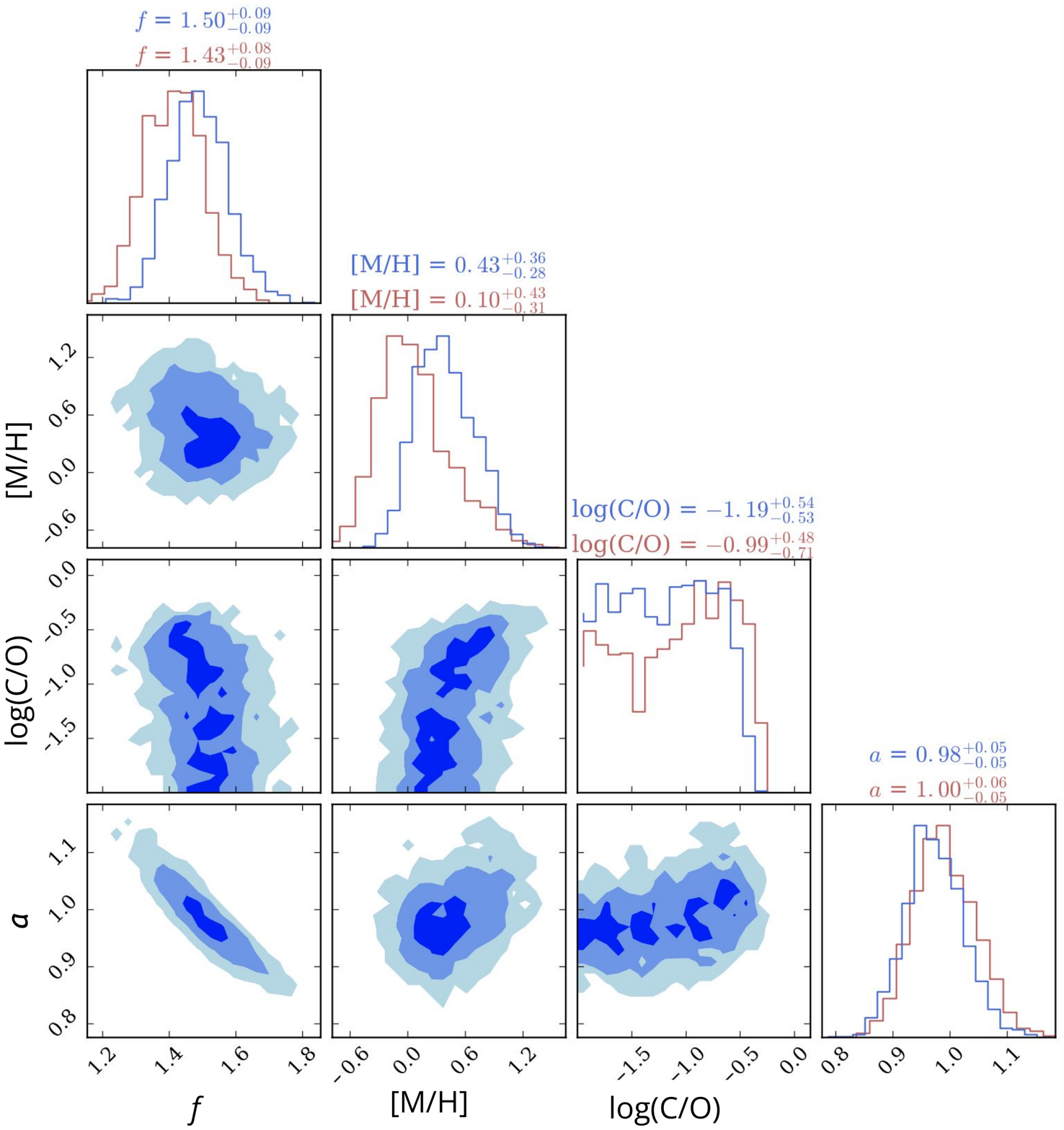}
    \caption{Posterior distributions of free parameters in the 1D Sc-CHIMERA model grid fit for the fit to the full data set (blue) and the fit to the data with the bluest 4 points removed (red). For the full data set, 2D histograms for pairs of parameters are shown in off-diagonal plots with 1, 2, and 3$\sigma$ regions shaded in light, medium, and dark blue, respectively. Histograms on the diagonal show 1D posterior probability distributions for each individual parameter.} 
    \label{fig:1dcorners}
\end{figure}

Figure~\ref{fig:1dspec} shows the resulting spectrum and T-P profile for the grid fit, and Figure~\ref{fig:1dcorners} shows a corner plot for the full posterior. The best grid fit had a reduced chi squared value of $\chi^{2}_{\nu}=1.24$ and showed a non-inverted T-P profile. The value of $f=1.50\pm0.09$ retrieved from the fit is consistent with 3D models of cloud-free hot Jupiters at the temperature of WASP-77Ab \citep{Parmentier2020}. The retrieved value of the scale factor $a$ was close to 1, which is consistent with the constraint on $f$ because with more heat redistribution we'd expect a less pronounced hotspot. The best fit metallicity was [M/H]$=0.43^{+0.36}_{-0.28}$. We note that this metallicity is significantly higher and less precise than the value derived from recent high-resolution observations - see Section~\ref{sec:highres} for a full discussion of these differences. The carbon-to-oxygen ratio is not well constrained, as we do not observe any resolved features of carbon-bearing molecules. However, the fit provides a $2\sigma$ upper limit of C/O$=0.78$, indicating that the planet likely has a solar or sub-solar C/O.

\section{Discussion}
\label{sec:discuss}

\subsection{\textit{HST} Water Feature Strength}

\begin{figure*}[h!]
    \centering
    \includegraphics[width=0.85\linewidth]{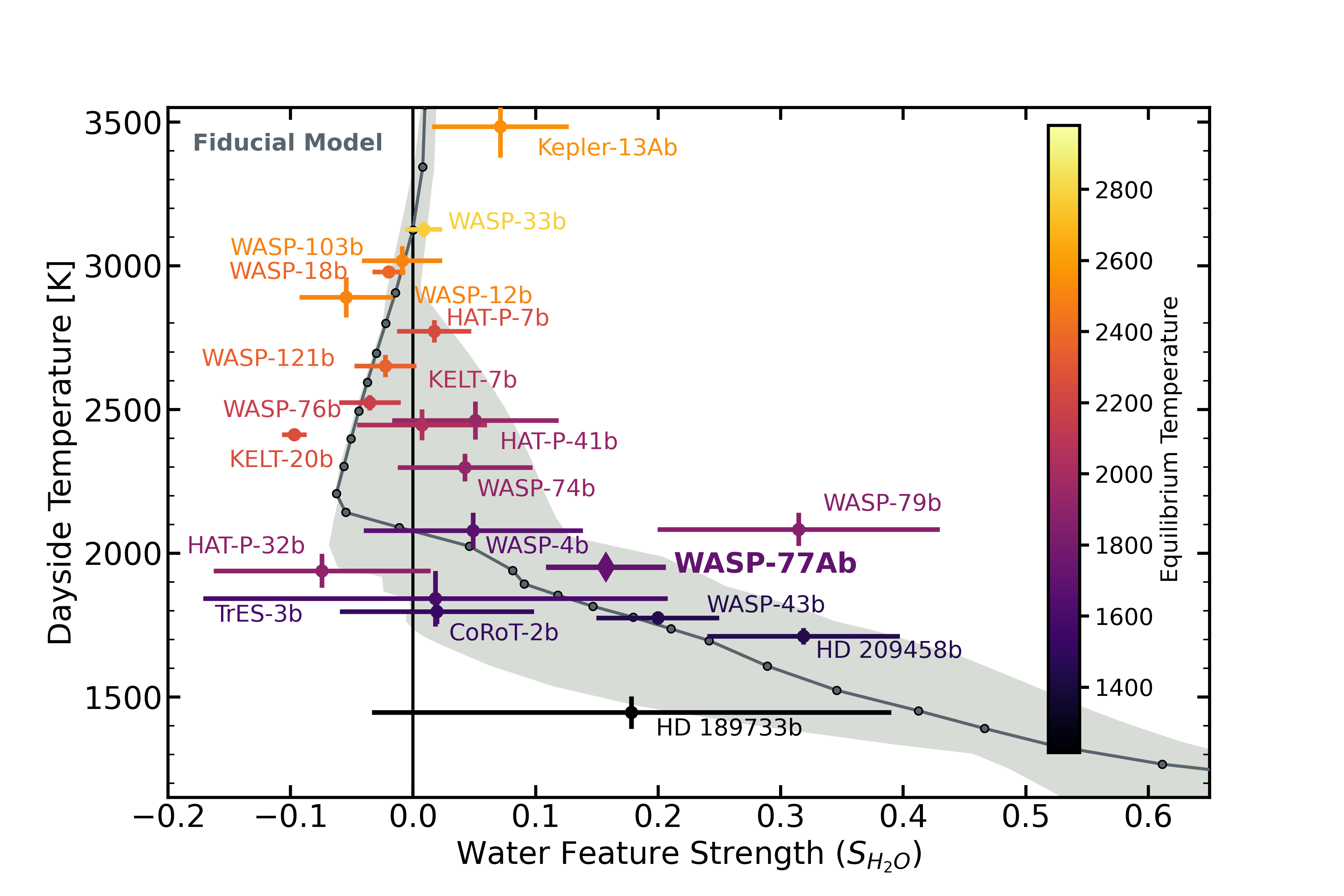}
    \caption{\textit{HST} water feature strength ($S_{H_{2}O}$) and dayside temperature ($T_{day}$) of WASP-77Ab (bold, diamond point) in the context of all other planets observed between $1.1-1.7$~$\mu$m with \textit{HST}/WFC3 \citep[colored, circular points;][]{Mansfield2021,Fu2022}. Both values are calculated following the descriptions in \citet{Mansfield2021}. Positive/negative values of $S_{H_{2}O}$ indicate features observed in absorption/emission, respectively, and a value of $S_{H_{2}O}=0$ indicates a featureless, blackbody-like spectrum. The grey points and shaded region show predictions from the 1D model grid presented in \citet{Mansfield2021}. WASP-77Ab has $S_{H_{2}O}=0.157 \pm 0.049$, indicating the presence of a strong water absorption feature in its spectrum. This value agrees with previously observed trends that planets below $T_{day}\approx2100$~K tend to have absorption features due to non-inverted T-P profiles, but that the scatter in water feature strengths for planets at similar temperatures suggests compositional differences in their atmospheres \citep{Fortney2008,Parmentier2018,Mansfield2021}.}
    \label{fig:colormag}
\end{figure*}

In order to place our observations of WASP-77Ab in the broader context of previous hot Jupiter secondary eclipse observations, we compared the observed water feature strength and derived metallicity to \textit{HST}/WFC3 observations of other hot Jupiters. We computed the \textit{HST} water feature strength $S_{H_{2}O}$ for WASP-77Ab following Equation~1 in \citet{Mansfield2021}. The water feature strength for WASP-77Ab is shown in Figure~\ref{fig:colormag} compared to the feature strengths for the data and models presented in \citet{Mansfield2021}. We find that the water feature strength of WASP-77Ab fits the previously observed trend, and matches the expectations from the self-consistent models of \citet{Mansfield2021}. Additionally, the fact that this planet shows a water feature in absorption at a dayside temperature of $\approx1900$~K disfavors models with high $\mathrm{C/O}\gtrsim0.7$, low metallicity $\mathrm{[M/H]}\lesssim-1.0$, or an amount of internal heating following \citet{Thorngren2019}, as such models predict a transition to inverted atmospheres below this temperature.

\subsection{Comparison to Gemini-S/IGRINS Results}
\label{sec:highres}

\begin{figure}[h!]
    \centering
    \includegraphics[width=\linewidth]{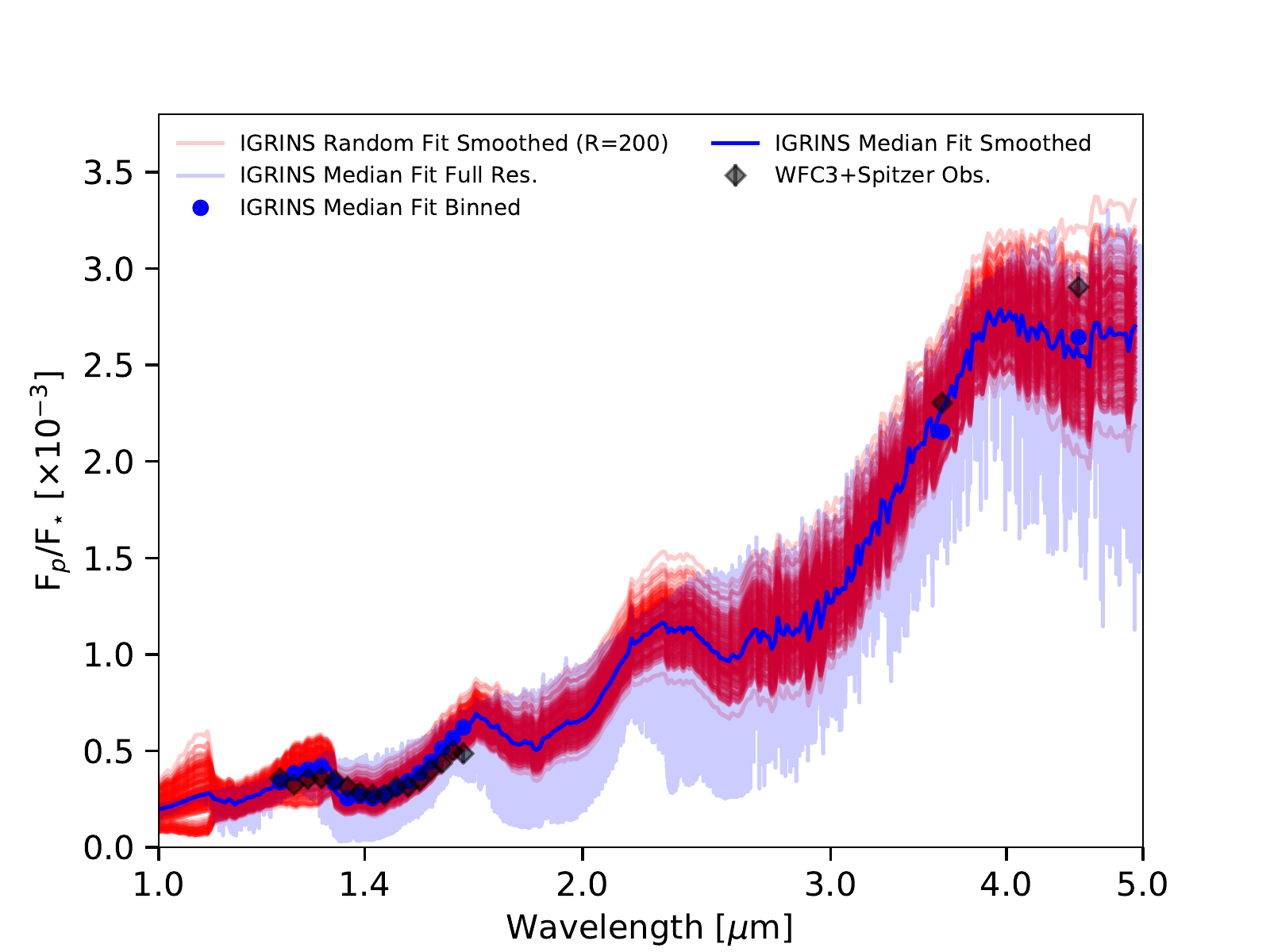}
    \caption{Comparison of our \textit{HST}/WFC3 and \textit{Spitzer} emission spectrum of WASP-77Ab (black points) with a fit of dayside emission from high-resolution Gemini-S/IGRINS observation \citep{Line2021}. The blue line and points show the median model fit to the IGRINS data, at full resolution and smoothed to the resolution of the WFC3 data, respectively. Red lines show 500 random draws from the posterior of the high-resolution fit smoothed to an R=100.}
    \label{fig:IGRINScomp}
\end{figure}

Confidence in composition and thermal structure inferences is bolstered when independent observations with different instruments arrive at the same conclusions. WASP-77Ab was recently observed near secondary eclipse at high resolution using the IGRINS spectrograph (R$\sim$45,000) on Gemini-South \citep{Line2021}. These observations spanned a wavelength range of $1.45-2.55$~$\mu$m, which allowed them to precisely constrain abundances of both water and carbon monoxide. Figure~\ref{fig:IGRINScomp} compares our WFC3 spectrum to the best-fit model from a high-resolution cross-correlation retrieval on these recent IGRINS observations. This plot also shows an ensemble of 500 spectra reconstructed from parameters drawn from the posterior probability distribution of that retrieval.

Figure~\ref{fig:IGRINScomp} shows that the extrapolated IGRINS model spectra are remarkably consistent with our WFC3 spectrum, providing cross-validation for these ground- and space-based observations. However, the best fit to the high-resolution observations retrieved a metallicity of [M/H]$=-0.48^{+0.15}_{-0.13}$ and a carbon-to-oxygen ratio of C/O$=0.59 \pm 0.08$ \citep{Line2021}. While their metallicity is consistent with the lower limit from our free retrieval (which is the same retrieval paradigm used in \cite{Line2021}), it is inconsistent with the metallicity we derive from the grid fit at $1.8\sigma$.

\begin{figure}[h!]
    \centering
    \includegraphics[width=\linewidth]{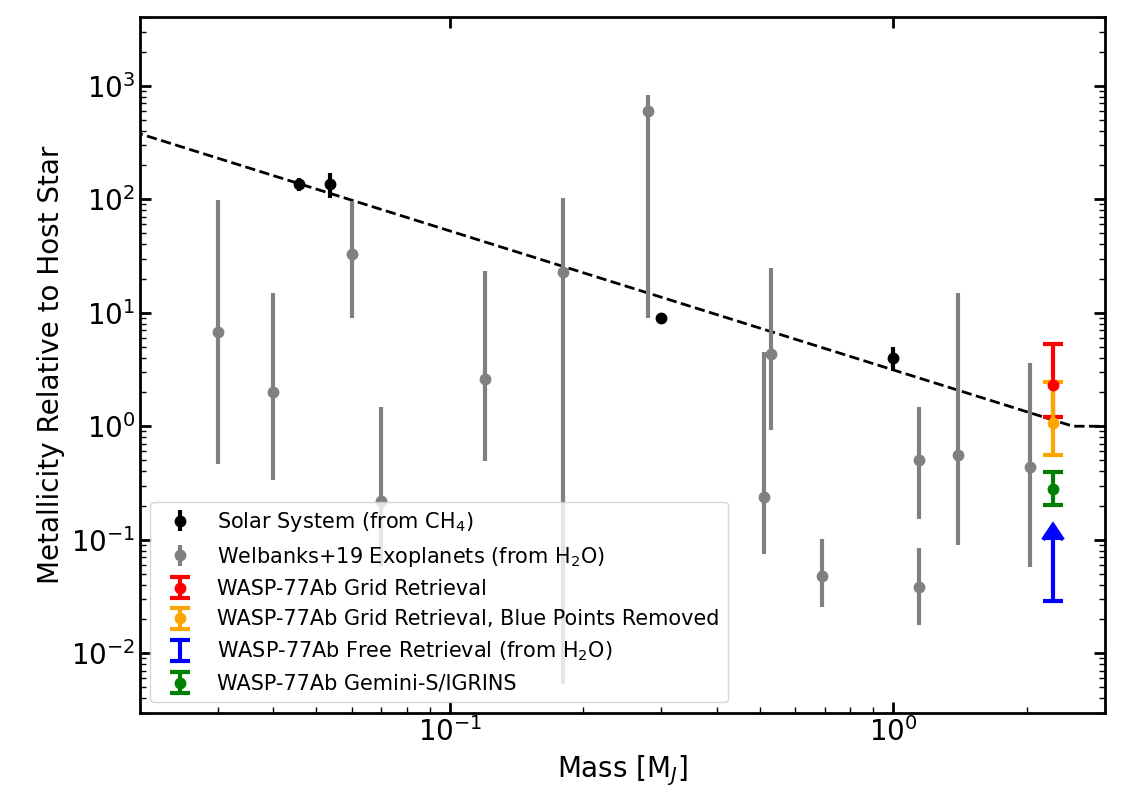}
    \caption{Atmospheric metallicity as a function of planet mass. Black points show solar system planet metallicities, which are based on measurements of [CH$_{4}$/H] \citep{Wong2004,Fletcher2009,Karkoschka2011,Sromovsky2011}. The black dashed line shows a fit to the solar system trend, but plateauing at 1 when the planet metallicity equals the stellar metallicity. Grey points show [H$_{2}$O/H] for previously observed exoplanets \citep{Welbanks2019}. We additionally compare four measurements of the metallicity of WASP-77Ab: [M/H] from our grid retrieval described in Section~\ref{sec:1dmodel} (red), [M/H] from the grid retrieval on the data with the bluest four points removed (orange, see Section~\ref{sec:highres}), [H$_{2}$O/H] from our free retrieval described in Section~\ref{sec:1dfree} (blue, lower limit only), and [C+O/H] from recent high-resolution observations with Gemini-S/IGRINS \citep[green,][]{Line2021}.}
    \label{fig:massmetal}
\end{figure}

Figure~\ref{fig:massmetal} shows the metallicities of WASP-77Ab derived from the free and grid retrievals compared to the IGRINS result. We investigated what could be driving the discrepancy in derived metallicities between our low-resolution WFC3 and \textit{Spitzer} data and the high-resolution IGRINS data and found that the higher metallicity we derive with our grid fits is driven by the strong downward slope of the bluest points in the WFC3 spectrum. We analyzed the two WFC3 visits independently and found that the downward slope at the blue end of the spectrum is consistent across both visits. We performed a grid fit to the WFC3+\textit{Spitzer} data with the bluest four points removed and derived a metallicity of [M/H]$=0.10^{+0.43}_{-0.31}$, which is more consistent with the high-resolution measurement. The results of this fit are shown in Figure~\ref{fig:1dcorners}. This result may indicate that the bluest part of the spectrum, which is not well fit by our equilibrium chemistry models, is influenced by disequilibrium chemistry. However, the lack of precise abundance measurements from our free retrieval demonstrates the difficulty of constraining chemistry in a non-equilibrium model with only low-resolution, low wavelength coverage data. Alternatively, this discrepancy may just be due to the sensitivity of low-resolution retrieval results to slight changes in the spectral shape. We note that our investigation here is not intended to provide a more accurate or precise metallicity measurement than that derived from the IGRINS observations, but rather to use a comparison of these two data sets to bolster confidence in the high-resolution result and discuss the limitations of deriving abundance constraints from low-resolution data alone. Additionally, techniques for extracting abundance measurements from high-resolution data are relatively new and have only been applied to a couple of data sets, so a comparison to the low-resolution results we present here is useful for assessing the validity of the high-resolution results, even if the low-resolution composition measurements are less well constrained.

We leave a joint retrieval combining our low-resolution \textit{HST} and \textit{Spitzer} data and the high-resolution Gemini-S/IGRINS data for a future paper (Smith et al.\ in prep.). Although it is outside the scope of this paper, such combined high-resolution and low-resolution fits can constrain the atmospheric composition even more tightly than either data set alone \citep{Brogi2019}. Additionally, in the near future, (\textit{JWST}) will measured the dayside emission spectrum of WASP-77Ab from 2.87 to 5.10\,$\mu$m (GTO 1274; PI Lunine). These data will further illuminate the atmospheric composition of WASP-77Ab.

\begin{acknowledgments}
This work was based on observations made with the NASA/ESA \textit{Hubble Space Telescope} that were obtained from the data archive at the Space Telescope Science Institute (STScI), which is operated by the Association of Universities for Research in Astronomy, Inc. under NASA contract NAS 5-26555. This work also used observations made with the \textit{Spitzer Space Telescope}, which is operated by the Jet Propulsion Laboratory, California Institute of Technology under a contract with NASA. We acknowledge an anonymous reviewer for helpful comments which improved the manuscript. Support for this work was provided by NASA through the NASA Hubble Fellowship grant HST-HF2-51485.001-A awarded by STScI. M. Mansfield acknowledges support from a NASA FINESST grant. J.M.D acknowledges support from the Amsterdam Academic Alliance (AAA) Program, and the European Research Council (ERC) European Union’s Horizon 2020 research and innovation programme (grant agreement no. 679633; Exo-Atmos). This work is part of the research programme VIDI New Frontiers in Exoplanetary Climatology with project number 614.001.601, which is (partly) financed by the Dutch Research Council (NWO).
\end{acknowledgments}

\clearpage 
\vspace{5mm}
\facilities{HST(WFC3), Spitzer(IRAC)}

\software{batman \citep{Kreidberg2015}, emcee \citep{Foreman2013}, matplotlib \citep{Hunter2007}, numpy \citep{vanderWalt2011}, pymultinest \citep{pymultinest}, pysynphot \citep{pysynphot2013}, scipy \citep{Virtanen2019}}




\begin{thebibliography}{}
\expandafter\ifx\csname natexlab\endcsname\relax\def\natexlab#1{#1}\fi
\providecommand{\url}[1]{\href{#1}{#1}}
\providecommand{\dodoi}[1]{doi:~\href{http://doi.org/#1}{\nolinkurl{#1}}}
\providecommand{\doeprint}[1]{\href{http://ascl.net/#1}{\nolinkurl{http://ascl.net/#1}}}
\providecommand{\doarXiv}[1]{\href{https://arxiv.org/abs/#1}{\nolinkurl{https://arxiv.org/abs/#1}}}

\bibitem[{{Ali-Dib}(2017)}]{AliDib2017}
{Ali-Dib}, M. 2017, MNRAS, 467, 2845, \dodoi{10.1093/mnras/stx260}

\bibitem[{{Arcangeli} {et~al.}(2018){Arcangeli}, {D{\'e}sert}, {Line}, {Bean},
  {Parmentier}, {Stevenson}, {Kreidberg}, {Fortney}, {Mansfield}, \&
  {Showman}}]{Arcangeli2018}
{Arcangeli}, J., {D{\'e}sert}, J.-M., {Line}, M.~R., {et~al.} 2018, ApJL, 855,
  L30, \dodoi{10.3847/2041-8213/aab272}

\bibitem[{{Beatty} {et~al.}(2017){Beatty}, {Madhusudhan}, {Tsiaras}, {Zhao},
  {Gilliland}, {Knutson}, {Shporer}, \& {Wright}}]{Beatty2017}
{Beatty}, T.~G., {Madhusudhan}, N., {Tsiaras}, A., {et~al.} 2017, AJ, 154, 158,
  \dodoi{10.3847/1538-3881/aa899b}

\bibitem[{{Bell} {et~al.}(2021){Bell}, {Dang}, {Cowan}, {Bean}, {D{\'e}sert},
  {Fortney}, {Keating}, {Kempton}, {Kreidberg}, {Line}, {Mansfield},
  {Parmentier}, {Stevenson}, {Swain}, \& {Zellem}}]{Bell2021}
{Bell}, T.~J., {Dang}, L., {Cowan}, N.~B., {et~al.} 2021, \mnras, 504, 3316,
  \dodoi{10.1093/mnras/stab1027}

\bibitem[{{Berta} {et~al.}(2012){Berta}, {Charbonneau}, {D{\'e}sert},
  {Miller-Ricci Kempton}, {McCullough}, {Burke}, {Fortney}, {Irwin}, {Nutzman},
  \& {Homeier}}]{Berta2012}
{Berta}, Z.~K., {Charbonneau}, D., {D{\'e}sert}, J.-M., {et~al.} 2012, ApJ,
  747, 35, \dodoi{10.1088/0004-637X/747/1/35}

\bibitem[{{Brogi} \& {Line}(2019)}]{Brogi2019}
{Brogi}, M., \& {Line}, M.~R. 2019, AJ, 157, 114,
  \dodoi{10.3847/1538-3881/aaffd3}

\bibitem[{{Buchner}(2016)}]{pymultinest}
{Buchner}, J. 2016, {PyMultiNest: Python interface for MultiNest}.
\newblock \doeprint{1606.005}

\bibitem[{{Cubillos} {et~al.}(2013){Cubillos}, {Harrington}, {Madhusudhan},
  {Stevenson}, {Hardy}, {Blecic}, {Anderson}, {Hardin}, \&
  {Campo}}]{Cubillos2013}
{Cubillos}, P., {Harrington}, J., {Madhusudhan}, N., {et~al.} 2013, \apj, 768,
  42, \dodoi{10.1088/0004-637X/768/1/42}

\bibitem[{{Edwards} {et~al.}(2020){Edwards}, {Changeat}, {Baeyens}, {Tsiaras},
  {Al-Refaie}, {Taylor}, {Yip}, {Bieger}, {Blain}, {Gressier}, {Guilluy},
  {Jaziri}, {Kiefer}, {Modirrousta-Galian}, {Morvan}, {Mugnai}, {Pluriel},
  {Poveda}, {Skaf}, {Whiteford}, {Wright}, {Zingales}, {Charnay}, {Drossart},
  {Leconte}, {Venot}, {Waldmann}, \& {Beaulieu}}]{Edwards2020}
{Edwards}, B., {Changeat}, Q., {Baeyens}, R., {et~al.} 2020, AJ, 160, 8,
  \dodoi{10.3847/1538-3881/ab9225}

\bibitem[{{Espinoza} {et~al.}(2017){Espinoza}, {Fortney}, {Miguel},
  {Thorngren}, \& {Murray-Clay}}]{Espinoza2017}
{Espinoza}, N., {Fortney}, J.~J., {Miguel}, Y., {Thorngren}, D., \&
  {Murray-Clay}, R. 2017, \apjl, 838, L9, \dodoi{10.3847/2041-8213/aa65ca}

\bibitem[{{Evans} {et~al.}(2017){Evans}, {Sing}, {Kataria}, {Goyal}, {Nikolov},
  {Wakeford}, {Deming}, {Marley}, {Amundsen}, {Ballester}, {Barstow},
  {Ben-Jaffel}, {Bourrier}, {Buchhave}, {Cohen}, {Ehrenreich}, {Garc{\'\i}a
  Mu{\~n}oz}, {Henry}, {Knutson}, {Lavvas}, {Lecavelier Des Etangs}, {Lewis},
  {L{\'o}pez-Morales}, {Mandell}, {Sanz-Forcada}, {Tremblin}, \&
  {Lupu}}]{evans17}
{Evans}, T.~M., {Sing}, D.~K., {Kataria}, T., {et~al.} 2017, \nat, 548, 58,
  \dodoi{10.1038/nature23266}

\bibitem[{{Feroz} {et~al.}(2009){Feroz}, {Hobson}, \& {Bridges}}]{multinest}
{Feroz}, F., {Hobson}, M.~P., \& {Bridges}, M. 2009, \mnras, 398, 1601,
  \dodoi{10.1111/j.1365-2966.2009.14548.x}

\bibitem[{{Fletcher} {et~al.}(2009){Fletcher}, {Orton}, {Teanby}, {Irwin}, \&
  {Bjoraker}}]{Fletcher2009}
{Fletcher}, L.~N., {Orton}, G.~S., {Teanby}, N.~A., {Irwin}, P.~G.~J., \&
  {Bjoraker}, G.~L. 2009, \icarus, 199, 351,
  \dodoi{10.1016/j.icarus.2008.09.019}

\bibitem[{{Foreman-Mackey} {et~al.}(2013){Foreman-Mackey}, {Hogg}, {Lang}, \&
  {Goodman}}]{Foreman2013}
{Foreman-Mackey}, D., {Hogg}, D.~W., {Lang}, D., \& {Goodman}, J. 2013, PASP,
  125, 306, \dodoi{10.1086/670067}

\bibitem[{{Fortney} {et~al.}(2008){Fortney}, {Lodders}, {Marley}, \&
  {Freedman}}]{Fortney2008}
{Fortney}, J.~J., {Lodders}, K., {Marley}, M.~S., \& {Freedman}, R.~S. 2008,
  ApJ, 678, 1419, \dodoi{10.1086/528370}

\bibitem[{{Fortney} {et~al.}(2013){Fortney}, {Mordasini}, {Nettelmann},
  {Kempton}, {Greene}, \& {Zahnle}}]{Fortney2013}
{Fortney}, J.~J., {Mordasini}, C., {Nettelmann}, N., {et~al.} 2013, \apj, 775,
  80, \dodoi{10.1088/0004-637X/775/1/80}

\bibitem[{{Fu} {et~al.}(2021){Fu}, {Deming}, {Lothringer}, {Nikolov}, {Sing},
  {Kempton}, {Ih}, {Evans}, {Stevenson}, {Wakeford}, {Rodriguez}, {Eastman},
  {Stassun}, {Henry}, {L{\'o}pez-Morales}, {Lendl}, {Conti}, {Stockdale},
  {Collins}, {Kielkopf}, {Barstow}, {Sanz-Forcada}, {Ehrenreich}, {Bourrier},
  \& {dos Santos}}]{Fu2020}
{Fu}, G., {Deming}, D., {Lothringer}, J., {et~al.} 2021, \aj, 162, 108,
  \dodoi{10.3847/1538-3881/ac1200}

\bibitem[{{Fu} {et~al.}(2022){Fu}, {Sing}, {Lothringer}, {Deming}, {Ih},
  {Kempton}, {Malik}, {Komacek}, {Mansfield}, \& {Bean}}]{Fu2022}
{Fu}, G., {Sing}, D.~K., {Lothringer}, J.~D., {et~al.} 2022, \apjl, 925, L3,
  \dodoi{10.3847/2041-8213/ac4968}

\bibitem[{{Horne}(1986)}]{Horne1986}
{Horne}, K. 1986, PASP, 98, 609, \dodoi{10.1086/131801}

\bibitem[{{Hubeny} {et~al.}(2003){Hubeny}, {Burrows}, \&
  {Sudarsky}}]{Hubeny2003}
{Hubeny}, I., {Burrows}, A., \& {Sudarsky}, D. 2003, \apj, 594, 1011,
  \dodoi{10.1086/377080}

\bibitem[{Hunter(2007)}]{Hunter2007}
Hunter, J.~D. 2007, Computing In Science \& Engineering, 9, 90,
  \dodoi{10.1109/MCSE.2007.55}

\bibitem[{{Husser} {et~al.}(2013){Husser}, {Wende-von Berg}, {Dreizler},
  {Homeier}, {Reiners}, {Barman}, \& {Hauschildt}}]{Husser2013}
{Husser}, T.~O., {Wende-von Berg}, S., {Dreizler}, S., {et~al.} 2013, A\&A,
  553, A6, \dodoi{10.1051/0004-6361/201219058}

\bibitem[{{Karkoschka} \& {Tomasko}(2011)}]{Karkoschka2011}
{Karkoschka}, E., \& {Tomasko}, M.~G. 2011, \icarus, 211, 780,
  \dodoi{10.1016/j.icarus.2010.08.013}

\bibitem[{{Kitzmann} {et~al.}(2018){Kitzmann}, {Heng}, {Rimmer}, {Hoeijmakers},
  {Tsai}, {Malik}, {Lendl}, {Deitrick}, \& {Demory}}]{Kitzmann2018}
{Kitzmann}, D., {Heng}, K., {Rimmer}, P.~B., {et~al.} 2018, ApJ, 863, 183,
  \dodoi{10.3847/1538-4357/aace5a}

\bibitem[{{Kreidberg}(2015)}]{Kreidberg2015}
{Kreidberg}, L. 2015, PASP, 127, 1161, \dodoi{10.1086/683602}

\bibitem[{{Kreidberg} {et~al.}(2014{\natexlab{a}}){Kreidberg}, {Bean},
  {D{\'e}sert}, {Line}, {Fortney}, {Madhusudhan}, {Stevenson}, {Showman},
  {Charbonneau}, {McCullough}, {Seager}, {Burrows}, {Henry}, {Williamson},
  {Kataria}, \& {Homeier}}]{Kreidberg2014}
{Kreidberg}, L., {Bean}, J.~L., {D{\'e}sert}, J.-M., {et~al.}
  2014{\natexlab{a}}, ApJL, 793, L27, \dodoi{10.1088/2041-8205/793/2/L27}

\bibitem[{{Kreidberg} {et~al.}(2014{\natexlab{b}}){Kreidberg}, {Bean},
  {D{\'e}sert}, {Benneke}, {Deming}, {Stevenson}, {Seager}, {Berta-Thompson},
  {Seifahrt}, \& {Homeier}}]{Kreidberg2014a}
---. 2014{\natexlab{b}}, Nature, 505, 69, \dodoi{10.1038/nature12888}

\bibitem[{{Kreidberg} {et~al.}(2018){Kreidberg}, {Line}, {Parmentier},
  {Stevenson}, {Louden}, {Bonnefoy}, {Faherty}, {Henry}, {Williamson},
  {Stassun}, {Beatty}, {Bean}, {Fortney}, {Showman}, {D{\'e}sert}, \&
  {Arcangeli}}]{Kreidberg2018}
{Kreidberg}, L., {Line}, M.~R., {Parmentier}, V., {et~al.} 2018, AJ, 156, 17,
  \dodoi{10.3847/1538-3881/aac3df}

\bibitem[{{Lanotte} {et~al.}(2014){Lanotte}, {Gillon}, {Demory}, {Fortney},
  {Astudillo}, {Bonfils}, {Magain}, {Delfosse}, {Forveille}, {Lovis}, {Mayor},
  {Neves}, {Pepe}, {Queloz}, {Santos}, \& {Udry}}]{Lanotte2014}
{Lanotte}, A.~A., {Gillon}, M., {Demory}, B.~O., {et~al.} 2014, \aap, 572, A73,
  \dodoi{10.1051/0004-6361/201424373}

\bibitem[{{Line} {et~al.}(2013){Line}, {Wolf}, {Zhang}, {Knutson}, {Kammer},
  {Ellison}, {Deroo}, {Crisp}, \& {Yung}}]{Line2013}
{Line}, M.~R., {Wolf}, A.~S., {Zhang}, X., {et~al.} 2013, \apj, 775, 137,
  \dodoi{10.1088/0004-637X/775/2/137}

\bibitem[{{Line} {et~al.}(2016){Line}, {Stevenson}, {Bean}, {Desert},
  {Fortney}, {Kreidberg}, {Madhusudhan}, {Showman}, \&
  {Diamond-Lowe}}]{Line2016}
{Line}, M.~R., {Stevenson}, K.~B., {Bean}, J., {et~al.} 2016, AJ, 152, 203,
  \dodoi{10.3847/0004-6256/152/6/203}

\bibitem[{{Line} {et~al.}(2021){Line}, {Brogi}, {Bean}, {Gandhi}, {Zalesky},
  {Parmentier}, {Smith}, {Mace}, {Mansfield}, {Kempton}, {Fortney}, {Shkolnik},
  {Patience}, {Rauscher}, {D{\'e}sert}, \& {Wardenier}}]{Line2021}
{Line}, M.~R., {Brogi}, M., {Bean}, J.~L., {et~al.} 2021, arXiv e-prints,
  arXiv:2110.14821.
\newblock \doarXiv{2110.14821}

\bibitem[{{Lothringer} {et~al.}(2018){Lothringer}, {Barman}, \&
  {Koskinen}}]{Lothringer2018}
{Lothringer}, J.~D., {Barman}, T., \& {Koskinen}, T. 2018, ApJ, 866, 27,
  \dodoi{10.3847/1538-4357/aadd9e}

\bibitem[{{Madhusudhan} {et~al.}(2014){Madhusudhan}, {Amin}, \&
  {Kennedy}}]{Madhusudhan2014}
{Madhusudhan}, N., {Amin}, M.~A., \& {Kennedy}, G.~M. 2014, \apjl, 794, L12,
  \dodoi{10.1088/2041-8205/794/1/L12}

\bibitem[{{Madhusudhan} {et~al.}(2017){Madhusudhan}, {Bitsch}, {Johansen}, \&
  {Eriksson}}]{Madhusudhan2017}
{Madhusudhan}, N., {Bitsch}, B., {Johansen}, A., \& {Eriksson}, L. 2017, MNRAS,
  469, 4102, \dodoi{10.1093/mnras/stx1139}

\bibitem[{{Madhusudhan} \& {Seager}(2009)}]{madhu2009}
{Madhusudhan}, N., \& {Seager}, S. 2009, \apj, 707, 24,
  \dodoi{10.1088/0004-637X/707/1/24}

\bibitem[{{Mansfield} {et~al.}(2018){Mansfield}, {Bean}, {Line}, {Parmentier},
  {Kreidberg}, {D{\'e}sert}, {Fortney}, {Stevenson}, {Arcangeli}, \&
  {Dragomir}}]{Mansfield2018}
{Mansfield}, M., {Bean}, J.~L., {Line}, M.~R., {et~al.} 2018, AJ, 156, 10,
  \dodoi{10.3847/1538-3881/aac497}

\bibitem[{{Mansfield} {et~al.}(2021){Mansfield}, {Line}, {Bean}, {Fortney},
  {Parmentier}, {Wiser}, {Kempton}, {Gharib-Nezhad}, {Sing},
  {L{\'o}pez-Morales}, {Baxter}, {D{\'e}sert}, {Swain}, \&
  {Roudier}}]{Mansfield2021}
{Mansfield}, M., {Line}, M.~R., {Bean}, J.~L., {et~al.} 2021, Nature Astronomy,
  \dodoi{10.1038/s41550-021-01455-4}

\bibitem[{{Marocco} {et~al.}(2021){Marocco}, {Eisenhardt}, {Fowler},
  {Kirkpatrick}, {Meisner}, {Schlafly}, {Stanford}, {Garcia}, {Caselden},
  {Cushing}, {Cutri}, {Faherty}, {Gelino}, {Gonzalez}, {Jarrett}, {Koontz},
  {Mainzer}, {Marchese}, {Mobasher}, {Schlegel}, {Stern}, {Teplitz}, \&
  {Wright}}]{Marocco2021}
{Marocco}, F., {Eisenhardt}, P. R.~M., {Fowler}, J.~W., {et~al.} 2021, \apjs,
  253, 8, \dodoi{10.3847/1538-4365/abd805}

\bibitem[{{Maxted} {et~al.}(2013){Maxted}, {Anderson}, {Collier Cameron},
  {Doyle}, {Fumel}, {Gillon}, {Hellier}, {Jehin}, {Lendl}, {Pepe}, {Pollacco},
  {Queloz}, {S{\'e}gransan}, {Smalley}, {Southworth}, {Smith}, {Triaud},
  {Udry}, \& {West}}]{Maxted2013}
{Maxted}, P.~F.~L., {Anderson}, D.~R., {Collier Cameron}, A., {et~al.} 2013,
  \pasp, 125, 48, \dodoi{10.1086/669231}

\bibitem[{{May} \& {Stevenson}(2020)}]{May2020}
{May}, E.~M., \& {Stevenson}, K.~B. 2020, \aj, 160, 140,
  \dodoi{10.3847/1538-3881/aba833}

\bibitem[{{Mikal-Evans} {et~al.}(2020){Mikal-Evans}, {Sing}, {Kataria},
  {Wakeford}, {Mayne}, {Lewis}, {Barstow}, \& {Spake}}]{MikalEvans2020}
{Mikal-Evans}, T., {Sing}, D.~K., {Kataria}, T., {et~al.} 2020, MNRAS, 496,
  1638, \dodoi{10.1093/mnras/staa1628}

\bibitem[{{Mordasini} {et~al.}(2016){Mordasini}, {van Boekel}, {Molli{\`e}re},
  {Henning}, \& {Benneke}}]{Mordasini2016}
{Mordasini}, C., {van Boekel}, R., {Molli{\`e}re}, P., {Henning}, T., \&
  {Benneke}, B. 2016, ApJ, 832, 41, \dodoi{10.3847/0004-637X/832/1/41}

\bibitem[{{{\"O}berg} {et~al.}(2011){{\"O}berg}, {Murray-Clay}, \&
  {Bergin}}]{Oberg2011}
{{\"O}berg}, K.~I., {Murray-Clay}, R., \& {Bergin}, E.~A. 2011, \apjl, 743,
  L16, \dodoi{10.1088/2041-8205/743/1/L16}

\bibitem[{{Parmentier} {et~al.}(2021){Parmentier}, {Showman}, \&
  {Fortney}}]{Parmentier2020}
{Parmentier}, V., {Showman}, A.~P., \& {Fortney}, J.~J. 2021, \mnras, 501, 78,
  \dodoi{10.1093/mnras/staa3418}

\bibitem[{{Parmentier} {et~al.}(2018){Parmentier}, {Line}, {Bean}, {Mansfield},
  {Kreidberg}, {Lupu}, {Visscher}, {D{\'e}sert}, {Fortney}, {Deleuil},
  {Arcangeli}, {Showman}, \& {Marley}}]{Parmentier2018}
{Parmentier}, V., {Line}, M.~R., {Bean}, J.~L., {et~al.} 2018, A\&A, 617, A110,
  \dodoi{10.1051/0004-6361/201833059}

\bibitem[{{Schneider} \& {Bitsch}(2021)}]{Schneider2021}
{Schneider}, A.~D., \& {Bitsch}, B. 2021, \aap, 654, A71,
  \dodoi{10.1051/0004-6361/202039640}

\bibitem[{{Sromovsky} {et~al.}(2011){Sromovsky}, {Fry}, \&
  {Kim}}]{Sromovsky2011}
{Sromovsky}, L.~A., {Fry}, P.~M., \& {Kim}, J.~H. 2011, \icarus, 215, 292,
  \dodoi{10.1016/j.icarus.2011.06.024}

\bibitem[{{Stassun} {et~al.}(2017){Stassun}, {Collins}, \&
  {Gaudi}}]{Stassun2017}
{Stassun}, K.~G., {Collins}, K.~A., \& {Gaudi}, B.~S. 2017, AJ, 153, 136,
  \dodoi{10.3847/1538-3881/aa5df3}

\bibitem[{{Stevenson} {et~al.}(2014){Stevenson}, {Bean}, {Seifahrt},
  {D{\'e}sert}, {Madhusudhan}, {Bergmann}, {Kreidberg}, \&
  {Homeier}}]{Stevenson2014a}
{Stevenson}, K.~B., {Bean}, J.~L., {Seifahrt}, A., {et~al.} 2014, \aj, 147,
  161, \dodoi{10.1088/0004-6256/147/6/161}

\bibitem[{{Stevenson} {et~al.}(2012){Stevenson}, {Harrington}, {Fortney},
  {Loredo}, {Hardy}, {Nymeyer}, {Bowman}, {Cubillos}, {Bowman}, \&
  {Hardin}}]{Stevenson2012a}
{Stevenson}, K.~B., {Harrington}, J., {Fortney}, J.~J., {et~al.} 2012, \apj,
  754, 136, \dodoi{10.1088/0004-637X/754/2/136}

\bibitem[{{STScI Development Team}(2013)}]{pysynphot2013}
{STScI Development Team}. 2013, {pysynphot: Synthetic photometry software
  package}.
\newblock \doeprint{1303.023}

\bibitem[{{Taylor} {et~al.}(2021){Taylor}, {Parmentier}, {Line}, {Lee},
  {Irwin}, \& {Aigrain}}]{Taylor2020}
{Taylor}, J., {Parmentier}, V., {Line}, M.~R., {et~al.} 2021, \mnras, 506,
  1309, \dodoi{10.1093/mnras/stab1854}

\bibitem[{{Thorngren} {et~al.}(2019){Thorngren}, {Gao}, \&
  {Fortney}}]{Thorngren2019}
{Thorngren}, D., {Gao}, P., \& {Fortney}, J.~J. 2019, ApJL, 884, L6,
  \dodoi{10.3847/2041-8213/ab43d0}

\bibitem[{{Thorngren} {et~al.}(2016){Thorngren}, {Fortney}, {Murray-Clay}, \&
  {Lopez}}]{Thorngren2016}
{Thorngren}, D.~P., {Fortney}, J.~J., {Murray-Clay}, R.~A., \& {Lopez}, E.~D.
  2016, \apj, 831, 64, \dodoi{10.3847/0004-637X/831/1/64}

\bibitem[{{Turner} {et~al.}(2016){Turner}, {Pearson}, {Biddle}, {Smart},
  {Zellem}, {Teske}, {Hardegree-Ullman}, {Griffith}, {Leiter}, {Cates},
  {Nieberding}, {Smith}, {Thompson}, {Hofmann}, {Berube}, {Nguyen}, {Small},
  {Guvenen}, {Richardson}, {McGraw}, {Raphael}, {Crawford}, {Robertson},
  {Tombleson}, {Carleton}, {Towner}, {Walker-LaFollette}, {Hume}, {Watson},
  {Jones}, {Lichtenberger}, {Hoglund}, {Cook}, {Crossen}, {Jorgensen},
  {Romine}, {Thompson}, {Villegas}, {Wilson}, {Sanford}, {Taylor}, \&
  {Henz}}]{Turner2016}
{Turner}, J.~D., {Pearson}, K.~A., {Biddle}, L.~I., {et~al.} 2016, \mnras, 459,
  789, \dodoi{10.1093/mnras/stw574}

\bibitem[{{van der Walt} {et~al.}(2011){van der Walt}, {Colbert}, \&
  {Varoquaux}}]{vanderWalt2011}
{van der Walt}, S., {Colbert}, S.~C., \& {Varoquaux}, G. 2011, Computing in
  Science and Engineering, 13, 22, \dodoi{10.1109/MCSE.2011.37}

\bibitem[{{Venturini} {et~al.}(2016){Venturini}, {Alibert}, \&
  {Benz}}]{Venturini2016}
{Venturini}, J., {Alibert}, Y., \& {Benz}, W. 2016, \aap, 596, A90,
  \dodoi{10.1051/0004-6361/201628828}

\bibitem[{{Virtanen} {et~al.}(2019){Virtanen}, {Gommers}, {Oliphant},
  {Haberland}, {Reddy}, {Cournapeau}, {Burovski}, {Peterson}, {Weckesser},
  {Bright}, {van der Walt}, {Brett}, {Wilson}, {Jarrod Millman}, {Mayorov},
  {Nelson}, {Jones}, {Kern}, {Larson}, {Carey}, {Polat}, {Feng}, {Moore}, {Vand
  erPlas}, {Laxalde}, {Perktold}, {Cimrman}, {Henriksen}, {Quintero}, {Harris},
  {Archibald}, {Ribeiro}, {Pedregosa}, {van Mulbregt}, \&
  {Contributors}}]{Virtanen2019}
{Virtanen}, P., {Gommers}, R., {Oliphant}, T.~E., {et~al.} 2019, arXiv
  e-prints, arXiv:1907.10121.
\newblock \doarXiv{1907.10121}

\bibitem[{{Welbanks} {et~al.}(2019){Welbanks}, {Madhusudhan}, {Allard},
  {Hubeny}, {Spiegelman}, \& {Leininger}}]{Welbanks2019}
{Welbanks}, L., {Madhusudhan}, N., {Allard}, N.~F., {et~al.} 2019, \apjl, 887,
  L20, \dodoi{10.3847/2041-8213/ab5a89}

\bibitem[{{Wong} {et~al.}(2004){Wong}, {Mahaffy}, {Atreya}, {Niemann}, \&
  {Owen}}]{Wong2004}
{Wong}, M.~H., {Mahaffy}, P.~R., {Atreya}, S.~K., {Niemann}, H.~B., \& {Owen},
  T.~C. 2004, \icarus, 171, 153, \dodoi{10.1016/j.icarus.2004.04.010}

\end{thebibliography}
\end{document}